\documentclass[aps,prb,twocolumn,showpacs,preprintnumbers,superscriptaddress,amsmath,amssymb]{revtex4}

\usepackage{graphicx}
\usepackage{dcolumn}
\usepackage{bm}
\usepackage{epstopdf}
\usepackage{color}
\usepackage{natbib}
\usepackage{ulem}

\newcommand{\bscco}{Bi$_{2}$Sr$_{2}$CaCu$_{2}$O$_{8+\delta}$\,}

\begin{document}

\title{On-chip sensing of hotspots in superconducting terahertz emitters}

\author{Xianjing Zhou}
\affiliation{Center for Nanoscale Materials, Argonne National Laboratory, Lemont, Illinois 60439, USA}

\author{Xu Han}
\affiliation{Center for Nanoscale Materials, Argonne National Laboratory, Lemont, Illinois 60439, USA}

\author{Dieter Koelle}
\affiliation{Physikalisches Institut and Center for Quantum Science in LISA$^+$, Universit\"at T\"ubingen, D-72076 T\"ubingen, Germany}

\author{Reinhold Kleiner}
\affiliation{Physikalisches Institut and Center for Quantum Science in LISA$^+$, Universit\"at T\"ubingen, D-72076 T\"ubingen, Germany}

\author{Xufeng Zhang}
\email{xufeng.zhang@anl.gov}
\affiliation{Center for Nanoscale Materials, Argonne National Laboratory, Lemont, Illinois 60439, USA}

\author{Dafei Jin}
\email{djin@anl.gov}
\affiliation{Center for Nanoscale Materials, Argonne National Laboratory, Lemont, Illinois 60439, USA}

\date{\today}

\begin{abstract}
Intrinsic Josephson junctions in high-temperature superconductor \bscco(BSCCO) are known for their capability to emit high-power terahertz photons with widely tunable frequencies. Hotspots, as inhomogeneous temperature distributions across the junctions, are believed to play a critical role in synchronizing the gauge-invariant phase difference among the junctions, so as to achieve coherent strong emission. Previous optical imaging techniques have indirectly suggested that the hotspot temperature can go higher than the superconductor critical temperature. However, such optical approaches often disturb the local temperature profile and are too slow for device applications. In this paper, we demonstrate an on-chip \textit{in situ} sensing technique that can precisely quantify the local temperature profile. This is achieved by fabricating a series of micro ``sensor" junctions on top of an ``emitter" junction and measuring the critical current on the sensors versus the bias current applied to the emitter. This fully electronic on-chip design could enable efficient close-loop control of hotspots in BSCCO junctions and significantly enhance the functionality of superconducting terahertz emitters.
\end{abstract}

\maketitle
\pretolerance=8000 

The major challenge in today's terahertz (THz) technologies is the lack of a reliable solid-state emitter, especially in the 0.1 -- 1~THz regime.\cite{Tono07} THz emitters made of high-temperature superconductor \bscco(BSCCO) have recently attracted a lot of attention\cite{Welp13,Kakeya2016} (for a recent review, see Ref.\,\onlinecite{Kleiner2019}). The building blocks of a BSCCO THz emitter are the intrinsic Josephson junctions (IJJs), which naturally exist in a BSCCO single crystal.\citealp{Klei92} When an IJJ is biased with a voltage $V$, it generates an oscillating supercurrent across the junction, at the frequency defined by the ac Josephson relation, \(f=V/\Phi_0\), where $\Phi_0$ is the flux quantum. The frequency upper limit is determined by the superconductor gap energy. For a low-temperature superconductor, e.g., niobium, it is about 750~GHz. But for BSCCO, it can in principle go to 20\,THz. Since the first high critical temperature ($T_{\rm c}$) superconducting THz emitter was reported in 2007,\cite{Ozy07} considerable efforts have been made to study\cite{Lin08,Hu2009,Wang09,Wang10,Guenon10,Tsuji10,Yurgens11,Bense11,Asai11,Tsuji12b,Kake12,Li12,Gross12,Gross13,Kake15,Rudau15,Rudau2016,Zhang2019} and develop\cite{An13,Bense13b,Seki13,Ji14,Asai14,Tsuji14,Kita14,Kashiwagi15b,Kashiwagi15c,Zhou15,Zhou15b,Hao15,Elarabi2017,Elarabi2018,Sun2018,Kashiwagi2018} BSCCO THz emitters. A variety of applications such as THz imaging and spectroscopy based on BSCCO THz emitter have been demonstrated.\cite{Tsuji12,Kashiwagi14,Kashiwagi14b,Nakade2016,Sun2017}

\begin{figure*}[tb]
	\includegraphics[scale=0.8]{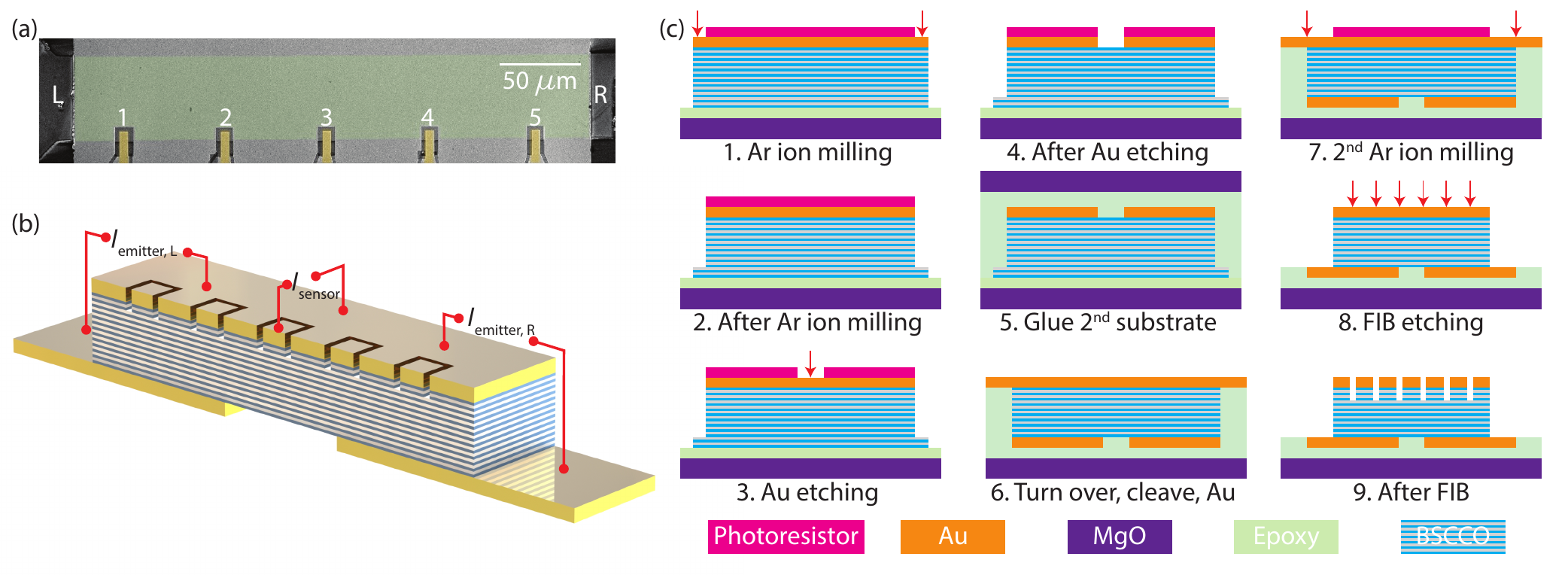}
	\caption{(a) Scanning electron microscope (SEM) picture of a stand-alone three-terminal superconducting terahertz emitter with five micro sensors located along one side of the top electrode. The emitter has two bottom electrodes labeled as L for the left and R for the right. The lateral dimension of the emitter (in green false color) is 300\,$\mu$m$\times$50\,$\mu$m and that of the sensors (in yellow false color) is about 5\,$\mu$m$\times$5\,$\mu$m. (b) Three-dimensional structure of the sample. The top electrode on the emitter is used as a common ground for the emitter and sensors. (c) Fabrication process for three-terminal terahertz emitter with on-chip sensors.}
	\label{fig:Sample}
\end{figure*}

Although the bias voltage can generate and tune the THz emission, the emission power is usually extremely low.\cite{Kleiner2007} An immediate approach to enhance the emission from a BSCCO THz emitter is to provide some resonant-feedback mechanism similar to a laser cavity. In the low-bias-current regime, when the target emission frequency from the ac Josephson relation matches the geometric resonance of the resonator (in most cases the junction stack itself), emission enhancement can indeed be achieved.\cite{Lin08,Hu2009} However, counterintuitively, the emission with narrowest linewidth is only detected in the high-bias-current regime, where the temperature distribution over the emitter becomes highly inhomogeneous and no apparent geometric resonance is involved. The local temperature ($T_{\rm{loc}}$) in some areas of the junction stack can even go above $T_{\rm{c}}$ and form hotspots.\cite{Wang09,Wang10,Guenon10,Yurgens11,Gross12} Although it has been believed that these hotspots help to synchronize the IJJ oscillations, there has been no definitive conclusion on how this happens in detail. Another unresolved puzzle is that the emission linewidth in the high-bias regime can be as narrow as 6\,MHz, which is at least 3 orders narrower than the GHz broadening in the low-bias regime.\cite{Li12,Gross13}

Due to the importance of hotspots in the high-bias strong emission regime, a variety of methods have been developed in order to image the inhomogeneous temperature distribution in BSCCO THz emitter.\cite{Wang09,Guenon10,Bense13c,Bense15,Minami14,Watanabe15,Kashiwagi2017} The existence of hotspots was first verified by low temperature scanning laser microscopy, which has a high spatial resolution limited only by the wavelength of the used laser. However, the laser imaging technique itself disturbs the system with local heating and the collected data requires sophisticated modeling and fitting.\cite{Gross12} Another approach is to use photoluminescence of europium chelate (EuTFC)\cite{Bense13c,Bense15} and SiC microcrystals\cite{Minami14, Watanabe15} to study the hotspots. However, this technique requires special preparation of the sample surface and has comparatively low spatial resolution. Thermoreflectance microscopy,\cite{Kashiwagi2017} which does not require the preparation of sample surface, cannot provide calibrated temperature profiles.

Past experiments indicate that hotspots favor locations where current is injected. In a previous work, we demonstrated a three-terminal device, in which we can choose the current injection from the left and the right bias electrodes.\cite{Zhou15b} With LTSLM, we showed that the location of hotspots can indeed be tuned by varying the current injection ratio between the electrodes. This in principle allows tuning the hotspots and thus controlling the THz emission by constructing a sensing-tuning feedback loop. However, the optical sensing technique has limited accuracy and speed, making feedback-loop control practically impossible. Here, we demonstrate a new, local temperature sensing method of BSCCO terahertz emitter. It does not rely on any extra components besides the BSCCO device itself. Micro BSCCO mesas with a few-layer junctions fabricated on top of the emitter act as local thermometers. Their maximum critical current $I_{\rm{c},\,max}$ directly indicates the local temperature. The feedback response time is only limited by the sensing electronics and can in principle reach a few nanoseconds.

\begin{figure*}[tb]
	\includegraphics[scale=1]{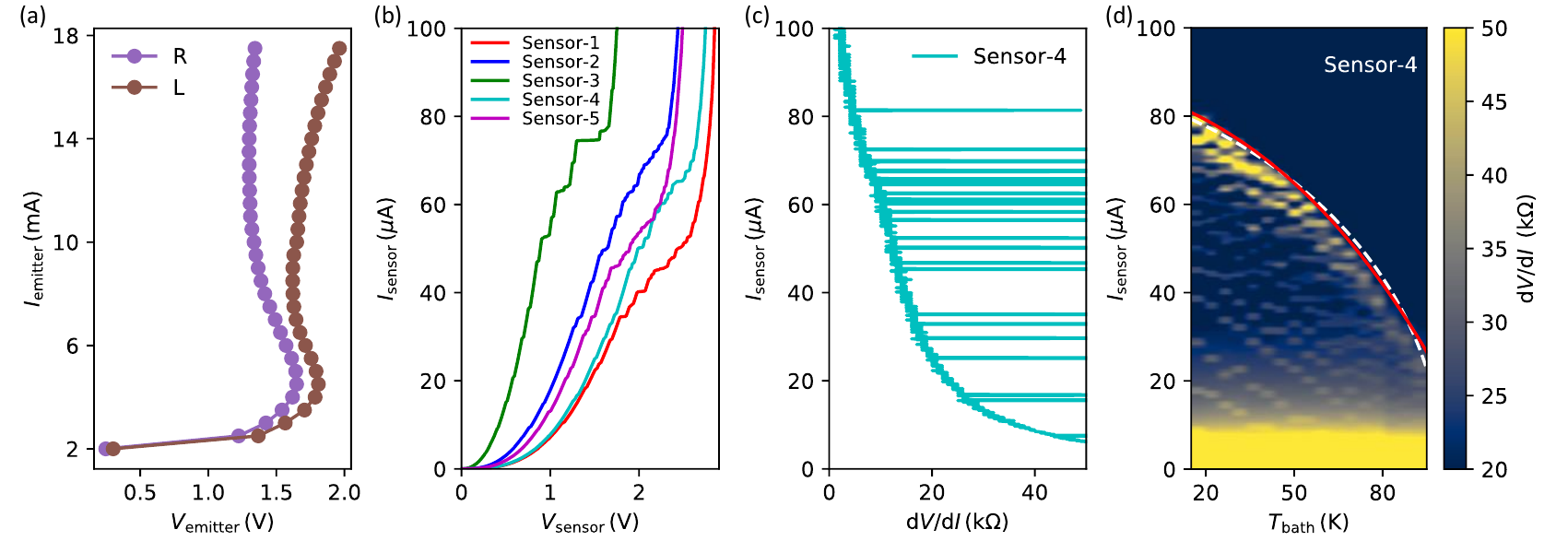}
	\caption{Measured properties of the emitter and sensors. (a) Current-voltage characteristics (IVCs) of the emitter in voltage state with bias current injected from the left electrode and the right electrode at bath temperature $T_{\rm{bath}}$\,=\,10\,K. (b) IVCs of sensors with $T_{\rm{bath}}$\,=\,10\,K. The steps shown in the IVCs are induced by a series of small junctions inside the sensors with different critical current, at which the superconducting state switches to the voltage state. The highest step defines the maximum critical current $I_{\rm{c,\,max}}$. (c) Differential resistance (d$V$/d$I$) of sensor-4, calculated from its IVC in (b). The steps can be seen more clearly. (d) Temperature and current dependence of d$V$/d$I$ of sensor-4. A clear boundary (approximately following the red line) indicates the decreasing $I_{\rm{c,\,max}}$ with increasing temperature. This means that $I_{\rm{c,\,max}}(T)$ can be used as a thermometric parameter to determine the local temperature on the emitter. Dashed white line shows the fitting curve to the function $j_{\rm{c}}(T)=j_{\rm{c}}(0)\sqrt{1-(T/T_{\rm{c}})^2}$. 
}	
	\label{fig:IVC}
\end{figure*}
In Fig.\,\ref{fig:Sample}(a), we show the scanning electron microscope (SEM) picture of a stand-alone three-terminal terahertz emitter with five micro sensors on it. We first fabricate the three-terminal terahertz emitter with the method reported previously.\cite{Zhou15b} This fabrication process for stand-alone three-terminal emitter is illustrated by step-1 to step-7 in Fig.\,\ref{fig:Sample}(c). The lateral dimension of the emitter is 300\,$\mu$m$\times$50\,$\mu$m, as marked in green false color, and the thickness is around 1\,$\mu$m, containing about 660 junction layers. We then fabricate small sensor junctions, as marked in yellow false color, on top of the emitter by focused ion beam (FIB) following the procedure step-8 to step-9 in Fig.\,\ref{fig:Sample}(c). In this sample, we fabricate five square sensors in total with side length around 5\,{$\mu$}m. The thickness of each sensor is about 30\,nm, containing about 20 junction layers, excluding the thickness of gold electrode. The lateral distance between two adjacent sensors is 60\,{$\mu$}m.

Experiment is performed in a Cryogenic Probe Station CRX-VF (Lake Shore Cryotronics, Inc.) with Keithley 4200A-SCS Parameter Analyzer. When measuring the local temperature, bias current of sensor $I_{\rm{sensor}}$ is applied between the top electrode of sensor and the top electrode of the emitter, as shown in the Fig.\,\ref{fig:Sample}(b). Thus $I_{\rm{sensor}}$ only goes through the sensor junctions and a few layers of the emitter, then reaches the top electrode of the emitter. With this configuration, the influence of $I_{\rm{sensor}}$ on the emitter can be neglected. The emitter is biased using two separated bottom electrodes, i.e., the left (L) and the right (R). The bias current can enter from L or R independently, through the emitter and reach the common top electrode.

\begin{figure*}[tb]
	\includegraphics[scale=1]{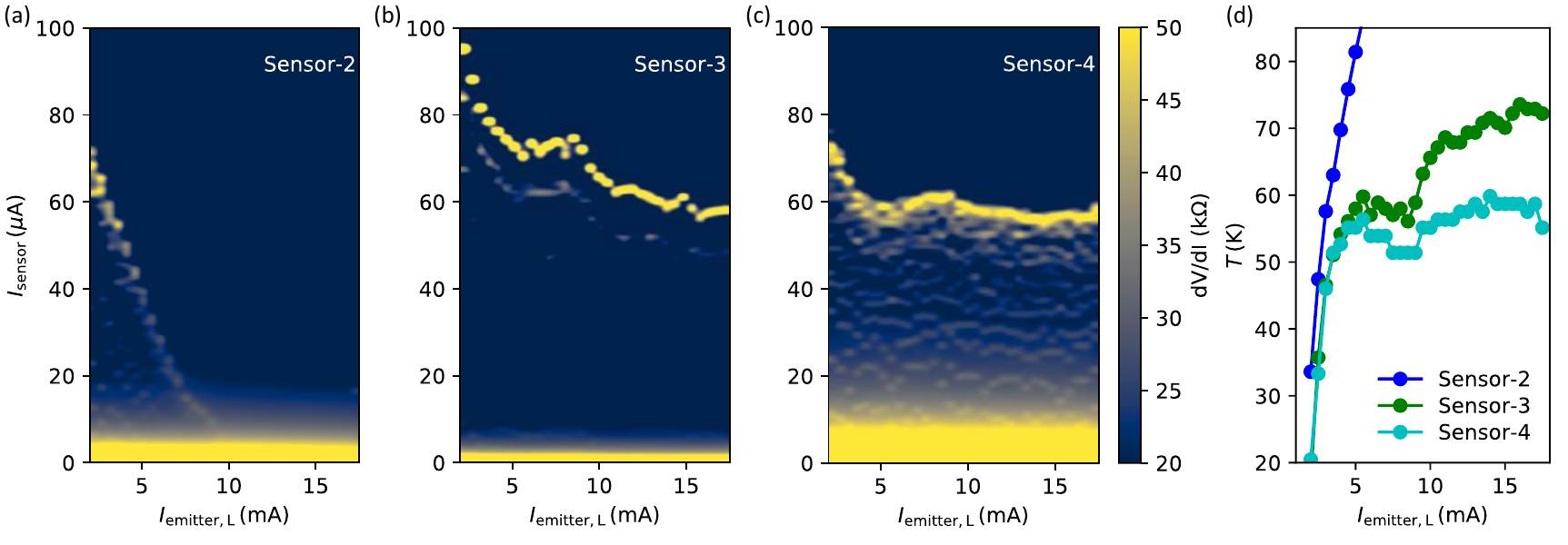}
	\caption{Properties of sensor-2, 3 and 4 when the emitter is current biased from left electrode. (a--c) Differential resistance (d$V$/d$I$) of senor-2, 3 and 4, versus the emitter current and sensor current. The dramatic difference in the boundary location indicates the different local temperature $T_{\rm{loc}}$ on the emitter. By comparing with the known $I_{\rm{c,\,max}}(T)$ (as in Fig.\ref{fig:IVC}(d)), we can accurately find the $T_{\rm{loc}}$ of each sensor. With increasing emitter current, the $T_{\rm{loc}}$ of sensor-2 raises up quickly above $T_{\rm{c}}$. By contrast, the $T_{\rm{loc}}$ of sensor-3 and 4 has only moderate increasing speed. We can observe nonmonotonic kinks in the curves of sensor-3 and 4; such features also show up in our theoretical simulation below using BSCCO's characteristic parameters. 
	}
	\label{fig:IC}
\end{figure*}

We first characterize the properties of the emitter and sensors. Figure\,\ref{fig:IVC}(a) gives the current voltage characterisitcs (IVCs) of emitter with current injected from different, L and R, electrodes at $T_{\rm{bath}}$\,=\,10\,K. The main reason for the difference in the IVCs is the hotspot formation at different locations, and the unavoidable small asymmetry of the sample. When the bias current of a three-terminal THz emitter is injected from different bottom electrodes, hotspots favor the location closer to the injecting port.\cite{Zhou15b}

Figure\,\ref{fig:IVC}(b) gives the IVCs of sensors at $T_{\rm{bath}}$\,=\,10\,K. The discrete steps in IVCs of sensors are induced by a series of junctions, with different $I_{\rm{c}}$ from different numbers of junction layers switching from the superconducting state to the voltage state. There is no further step once sensor current is above $I_{\rm{c,\,max}}$, i.e., all the junctions in the sensor are in the voltage state. Different sensors have slightly different $I_{\rm{c,\,max}}$ because of the small differences in the junction cross-sections. But for each given sensor, its $I_{\rm{c,\,max}}$ is always between 60 - 100 $\mu$A at $T_{\rm{bath}}$ \,=\,10\,K, corresponding to a critical current density $j_{\rm{c}}$ around 240 - 400\,A/cm$^2$. Switching steps can be seen more clearly in the differential resistance (d$V$/d$I$), e.g., that of sensor-4 in Fig.\,\ref{fig:IVC}(c). The joule heating induced by the sensor is estimated to be around 1\,nW and can be further reduced to 1\,pW with smaller junction size. Heating powers of these levels are negligible for our studies.

The relationship between $I_{\rm{c, max}}$ and temperature, for sensor-4 as an example, is shown in Fig.\,\ref{fig:IVC}(d), where $I_{\rm{c, max}}$ manifests as the boundary between the dark and bright areas (marked with a red line) . The color corresponds to the differential resistance d$V$/d$I$ as a function of $I_{\rm{sensor}}$ and $T_{\rm{bath}}$. The boundary $I_{\rm{c,\,max}}$-$T$ can be fitted by $j_{\rm{c}}(T)=j_{\rm{c}}(0)\sqrt{1-(T/T_{\rm{c}})^2}$, as shown by the white dashed line. As temperature goes up, $I_{\rm{c,\,max}}$ goes down. It is a monotonic function of temperature, and hence can be used as a thermometric parameter to indicate the local temperature. We perform the same measurement for each sensor to obtain its own $I_{\text{c,max}}$-$T$ relation. This eliminates any artifacts due to slight differences of sensor geometry or electrode contact.

Next we explain how to read out the local temperature on the emitter using the micro sensors (We focus on sensor-2, sensor-3 and sensor-4). We first vary the temperature distribution of the emitter by biasing current only from the electrode L and ramping its value. When the emitter is biased with a constant current, e.g., 2\,mA or 4\,mA, IVCs of the sensors are measured. d$V$/d$I$ of each sensor as a function of $I_{\rm{sensor}}$ and $I_{\rm{emitter}}$ are acquired, as shown in Fig.\,\ref{fig:IC}(a-c). The boundary between dark and bright area here is the relationship between $I_{\rm{c,\,max}}$ of each sensor and emitter current. Afterwards, the local temperature of the sensor as a function of emitter current can be calculated from the already obtained $I_{\rm{c,\,max}}$-$T$ curve of each sensor, e.g. of sensor-4 shown in Fig.\,\ref{fig:IVC}(d).

\begin{figure}[tb]
	\includegraphics[scale=1]{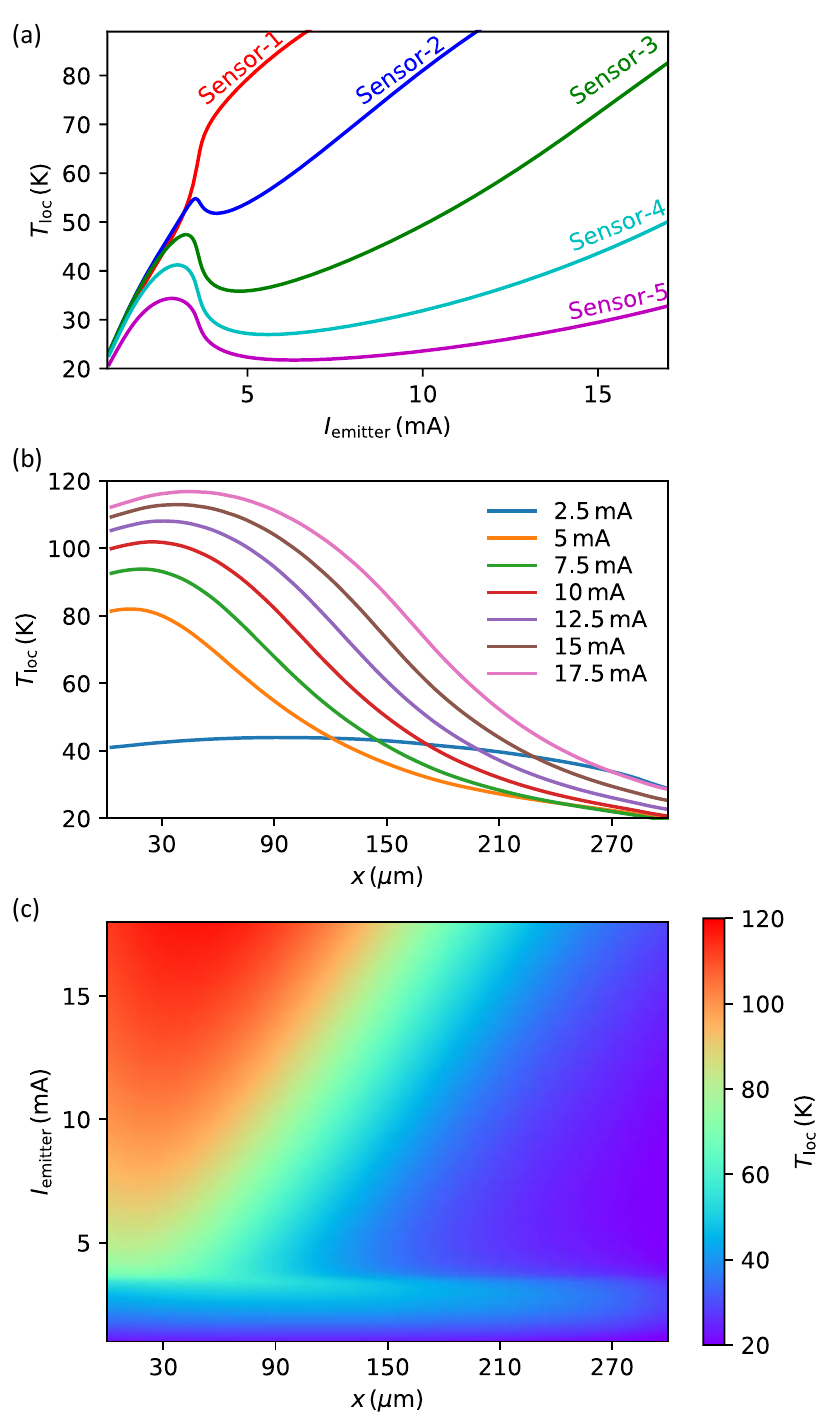}
	\caption{Theoretical simulation based on realistic Josephson junction electric circuit and heat transfer models of BSCCO with the same geometric configuration of the sample. (a) Local temperature $T_{\rm{loc}}$ at the position of each sensor as a function of emitter current. (b) Local temperature as a function of position for several selected emitter currents. (c) Color plot of the local temperature as a function of emitter current and the location along the line of sensor array.}	
	\label{fig:Simulation}
\end{figure}

As shown in Fig.\,\ref{fig:IC}(d), for a fixed emitter current, the local temperature of sensor-2 is always the highest and that of sensor-4 is always the lowest. We can also look at how the local temperature on the same sensor varies with the emitter current. When the emitter current increases from 2\,mA to 6\,mA, the local temperature of all sensors goes up. However, as the emitter current further increases from 6\,mA to 9\,mA, the local temperature of sensor-2 continues increasing beyond the range of calibration curve, while those of sensor-3 and sensor-4 decrease. When the emitter current goes above 9\,mA, the local temperature of sensor-3 and sensor-4 go up again. As elaborated below, this nonmonotonic feature agree well with the theoretically simulated curves which are determined by BSCCO's intrinsic thermal and electrical properties in the investigated temperature range.

From the IVCs of emitter in Fig.\,\ref{fig:IVC}(a), the current range 6\,mA to 9\,mA is the regime where a hotspot starts to form. Given the emitter current is injected from the left electrode, which is closer to sensor-2, we can infer that the hotspot starts to form at this current range and at a position near sensor-2 but far from sensor-4. Therefore, the boundary of hotspot should be located between sensor-2 and sensor-3. These results show that we can indeed sense the hotspot with these on-chip sensors. Moreover, this technique is general and scalable--the spatial sensing resolution can be improved by reducing the size and spacing of the sensors and increasing the number of sensors.

\begin{figure*}[tb]
	\includegraphics[scale=1.1]{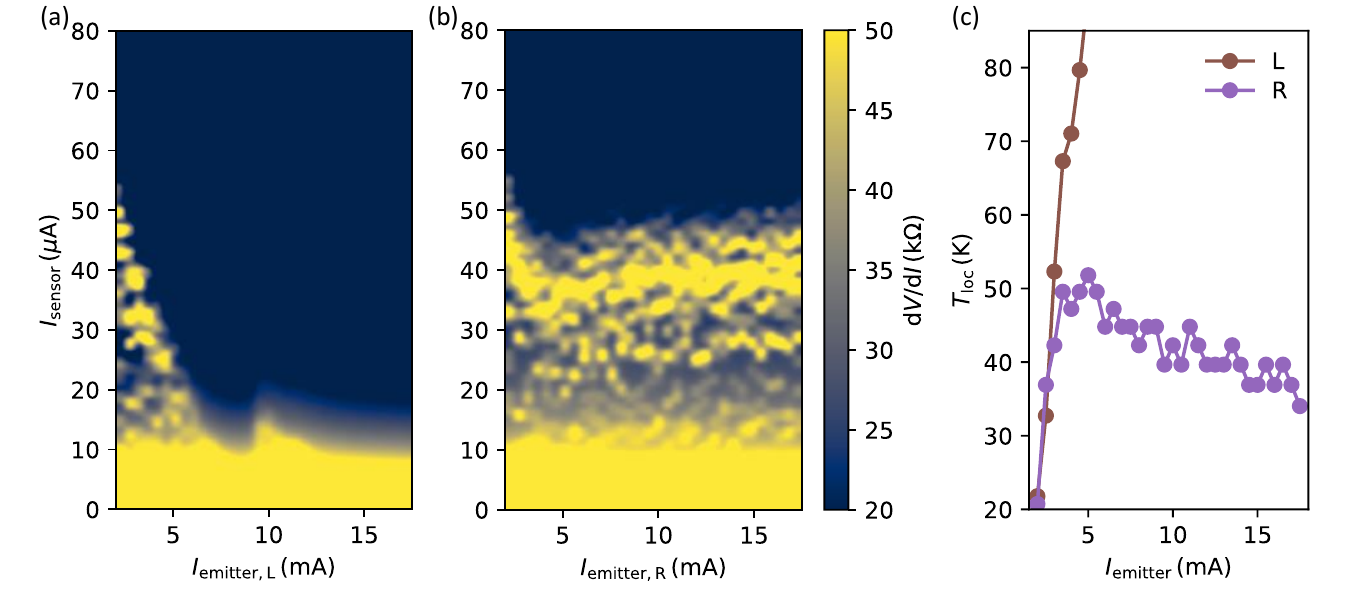}
	\caption{Comparison of the differential resistance of sensor-1 when the emitter current is injected from the left electrode (a) or the right electrode (b). For (a), the $T_{\rm{loc}}$ of sensor-1 increases rapidly, whereas for (b), it almost stays constant. 
}
	\label{fig:L&R}
\end{figure*}


In Fig.\,\ref{fig:Simulation}, we show the theoretically simulated hotspot formation and expansion by solving the nonlinear equation $-\nabla[\kappa(T(\textbf{r}))\nabla{T(\textbf{r})}]$=$\rho[T(\textbf{r})]j^2(\textbf{r})$, where is $\kappa$ the thermal conductivity of material and $\rho$ is the electrical resistivity of material. The simulation was done using COMSOL Multiphysics. All the geometric parameters in the simulation are extracted from the real sample, and the material parameters are the same as in Ref.\,\onlinecite{Gross12}. 

In the simulated results shown in Fig.\,\ref{fig:Simulation}(a), with the increasing emitter current from 1\,mA, the local temperature of sensor-1 through sensor-5 increases at first. But after that, there is a current regime where the local temperature of sensor-1 continues increasing but that of other sensors decrease. When the emitter current further increases, the temperature of all these sensors rises again. The formation and movement of hotspot can be seen more clearly in Fig.\,\ref{fig:Simulation}(b) where we plot the temperature profile along the edge of the emitter where all the sensors locate, for several selected emitter currents. In the 2D temperature plot in Fig.\,\ref{fig:Simulation}(c), as the bias current increases, the hotspot (in red color) begins to form while the local temperature of distant area starts to decrease. Then the boundary of hotspot moves from sensor-1 to sensor-5.

Besides demonstrating the formation and expansion of hotspot when emitter current is biased from electrode L, on-chip sensors can also reveal that the hotspot is formed at different locations when the current is injected from different electrodes. In Fig.\,\ref{fig:L&R}, we show the plots of d$V$/d$I$ and local temperature of sensor-1 when emitter current is injected from electrode L or electrode R at $T_{\rm{bath}}$\,=\,10\,K. When emitter is biased from electrode L, the local temperature of sensor-1 quickly goes above $T_{\rm{c}}$ as the emitter current increases. However, when the emitter is biased from the electrode R, the local temperature of sensor-1 first increases to around 50\,K and then goes back to around to 40\,K. We can infer from this result unambiguously that the hotspot locates near or far away from sensor-1 when the emitter current is injected from the left or the right electrode. This is consistent with our previous result.

The demonstrated functionality of these sensors will enable close-loop feedback control of the hotspots on BSCCO and hence the THz emission. For instance, the location of the maximum $T_{\rm{loc}}$ can be used as a feedback parameter to vary the ratio between the emitter current from the left electrode and the right electrode. The feedback response time is only limited by the sensing electronics and can in principle reach a few nanoseconds.

To summarize, we have demonstrated a novel approach for on-chip sensing of hotspots on BSCCO terahertz emitter. Since hotspots are directly associated with the underlying mechanism of the terahertz emission synchronization, our \textit{in situ} sensing technique can play a critical role in real-time control over the terahertz emission. The micro-mesa intrinsic Josephson-junction array demonstrated here provides a promising solution for developing next-generation terahertz circuits. Furthermore, they may open a new route for designing high-quality nonlinear superconducting devices, essential for quantum information science.

\begin{acknowledgements}

This work was performed in part at the Center for Nanoscale Materials, a U.S. Department of Energy Office of Science User Facility, and supported by the U.S. Department of Energy, Office of Science, under Contract No. DE-AC02-06CH11357. D.K. and R.K. acknowledge funding support by EU-FP6-COST Action CA16218 and the Deutsche Forschungsgemeinschaft via project KL93013/2. X.Z. and D.J. thank Daniel Lopez for scientific discussion and Brandon Fisher for technical assistance.

\end{acknowledgements}

\bibliography{BSCCO-Onchip-Sensing}

\end{document}